\documentclass{emulateapj}
\usepackage{color}
\usepackage{bm}
\usepackage{hyperref}
\usepackage{multirow}
\usepackage{lineno}
\hypersetup{
    colorlinks=true,
    linkcolor=blue,
    filecolor=magenta,      
    urlcolor=cyan,
    citecolor=blue,
    pdftitle={Overleaf Example},
    pdfpagemode=FullScreen,
    }

\slugcomment{}

\shorttitle{Synchrotron Polarization}

\begin{document}


\title{Polarization of synchrotron emission from rotation powered pulsars}


\author{Seima Saeki\altaffilmark{1}}
\email{d255401@hiroshima-u.ac.jp}
\author{Shota Kisaka\altaffilmark{1,2}}
\author{Jumpei Takata\altaffilmark{3}}


\altaffiltext{1}{Physics Program, Graduate School of Advanced Science and Engineering, Hiroshima University, 1-3-1, Kagamiyama, Higashi-Hiroshima, Hiroshima 739-8526, Japan}
\altaffiltext{2}{Hiroshima Astrophysical Science Center, Hiroshima University, 1-3-1, Kagamiyama, Higashi-Hiroshima, Hiroshima 739-8526, Japan}
\altaffiltext{3}{Department of Astronomy, School of Physics, Huazhong University of Science and Technology, Wuhan 430074, China}


\begin{abstract}
With the launch of the Imaging X-ray Polarimeter Explorer in 2021, recent observations have provided X-ray polarization data for several rotation-powered pulsars.
To fully exploit these data, it is important to better understand the relationship between polarization properties and the magnetospheric structure.
In this paper, we investigate polarization properties in the outer gap model, incorporating both outward emission and inward emission toward the rotation axis, which has not been considered previously.
We find that the polarization position angle varies at the intensity peaks, typically accompanied by a high polarization degree, as is in the case for outward emission.
Moreover, the variation in the polarization position angle tends to exceed $50^{\circ}$ when the intensity peaks of outward and inward emissions overlap, compared to cases with only outward emission.
We compare the polarization model including inward emission with IXPE observations. For example, for PSR B1509-58 where inward emission is proposed to be dominant in the outer gap model, we find that observed PA swing prefers a line of sight from the southern hemisphere rather than the northern hemisphere. 
\end{abstract}


\keywords{polarization -- radiation mechanisms: non-thermal -- pulsars: general}




\section{Introduction}
\label{sec:introduction}
Pulsed emissions are detected across a wide range of wavelengths from radio to gamma-ray from rotation powered pulsars \citep{2005AJ....129.1993M, 2023ApJ...958..191S}.
Accelerated particles by the strong electromagnetic field in the pulsar magnetospheres are thought to cause the emissions.
It remains unknown where and how particles are accelerated in the pulsar magnetosphere.
No magnetic attenuation caused by the strong magnetic field \citep{1966RvMP...38..626E} near the surface was detected in GeV emission \citep{2023ApJ...958..191S}, indicating that the gamma-ray emission region is far from the stars.
Some emission models have been proposed such as the outer gap model \citep{1986ApJ...300..500C} and the current sheet model \citep{2002A&A...388L..29K}.
These models reproduce some features of observed light curves \citep{2007ApJ...656.1044T, 2013MNRAS.434.2636P}. 
However, it remains challenging to distinguish between these models based solely on current observations, as they can reproduce similar features in the observed light curves.

Polarization observations provide additional constraints on the magnetic field geometry, emission mechanisms, and emission regions. 
The polarization degree (PD) and polarization position angle (PA) are key observables, probing the magnetic field ordering and tracing the orientation of the projected magnetic field, respectively. 
In pulsars, relativistic beaming allows emission from different regions of the magnetosphere to be observed at different rotational phases, making phase-resolved polarization observations particularly informative. 
Such observations have so far been limited largely to the Crab pulsar \citep{2023NatAs...7..602B, 2017NatSR...7.7816C, 2022MNRAS.512.2827L, 1996MNRAS.282.1354G, 2009MNRAS.397..103S}, except in the radio band.
Since the properties of pulsed emission strongly depend on the inclination angle between the rotation and magnetic axes, as well as the viewing geometry \citep{2004ApJ...606.1125D, 2010ApJ...715.1270B, 2017ApJ...840...73H}, obtaining polarization data for multiple pulsars is essential for a systematic understanding of magnetospheric structures.
In this context, the Imaging X-ray Polarimeter Explorer (IXPE) \citep{2022JATIS...8b6002W}, launched in 2021, has enabled polarization observations for several pulsars \citep{2023NatAs...7..602B, 2023ApJ...957...23R, 2024ApJ...962...92X, 2025A&A...699A..33B}, and further observations are expected to expand the available data set.

In this study, we focus on the outer gap model \citep{1986ApJ...300..500C}.
In the outer gap model, particle acceleration and high-energy emission occur in a finite region of the open-field-line zone beyond the null charge surface (NCS).
The NCS is defined as the location where the Goldreich-Julian (GJ) charge density $\rho_{\rm GJ}\equiv-\bm{\Omega}\cdot\bm{B}/(2\pi c)$ \citep{1969ApJ...157..869G}, becomes zero. 
Charged particles accelerated by the electric field emit gamma-ray photons tangentially to the local magnetic field lines through curvature emission. These photons produce electron-positron pairs through collision with X-ray photons from the stellar surface and/or magnetosphere.
The resulting electrons and positrons emit X-ray photons via synchrotron emission.
The outer gap model is consistent with the observed trend that most young pulsars exhibit single- or double-peaked gamma-ray light curves \citep{2010ApJS..187..460A, 2013ApJS..208...17A, 2023ApJ...958..191S, 2011MNRAS.415.1827T}.

Within the outer gap, charged particles are accelerated not only outward from the neutron star but also inward toward the stellar surface.
Gamma-ray photons emitted by these inward-moving particles are expected to be difficult to detect, as they are likely to undergo pair creation either through interactions with thermal X-ray photons from the neutron star (photon–photon pair production) or via magnetic pair creation in the strong magnetic field near the stellar surface.
However, the secondary particles produced in these interactions can emit synchrotron emission in the optical to X-ray and soft gamma-ray bands, allowing the inward emission component to be observed without significant attenuation.

Indeed, in the Vela pulsar, X-ray intensity peaks observed at rotational phases where no corresponding gamma-ray peaks are seen \citep{2002ApJ...576..376H} have been suggested to originate from inward emission \citep{2008MNRAS.386..748T, 2011ApJ...739...14K}.
In addition, for PSR B1509-58, whose spectrum peaks at softer energies than typical gamma-ray pulsars \citep{2010ApJ...714..927A}, inward emission has been proposed as a possible origin of both soft gamma-ray and X-ray emission within framework of the outer gap model \citep{2013ApJ...764...51W, 2014MNRAS.445..604W}.
These studies suggest that inward emission may contribute significantly in the X-ray band.
However, the polarization properties of such inward emission have not yet been systematically investigated.
With the increasing availability of X-ray polarization data from IXPE, understanding the polarization properties of inward emission is crucial for extracting information on the magnetospheric structure through comparison with observations.

In this study, we investigate the polarization characteristics of inward emission predicted by the outer gap model.
\S \ref{sec:emission model} describes the model assumptions and parameters.
\S \ref{sec:results} presents the polarization properties for outward emission, inward emission, and their overlap.
\S \ref{sec:discussion} compares the results with observational data and discusses thier implications.

\section{Model}
\label{sec:emission model}
In this study, the synchrotron emission model as pulsed emission from pulsar magnetospheres is considered.
To calculate the emission intensity, the PA, and the PD for each rotational phase, the magnetic field structure, and the emission region, the emission direction must be specified.
The calculation method used in this paper is almost the same as \citet{2007ApJ...656.1044T}.
While the same points are magnetic field structure of vacuum rotating dipole field, emission region of outer gap (\S \ref{subsec:magnetic field structure and emission region}) and polarization direction (\S \ref{subsec:stokes parameters}), the main difference lies in the emission direction. 
The emissions directed away from and toward the axis of rotation of the star are defined as the outward emission and the inward emission, respectively.
We considered not only outward emission but also inward emission (\S \ref{subsec:emission direction}) while only outward emission was taken into account in \citet{2007ApJ...656.1044T}.

\subsection{Magnetic Field Structure and Emission Region}
\label{subsec:magnetic field structure and emission region}
For simplicity, the vacuum rotating dipole field \citep{1955AnAp...18....1D, 1997PhDT.........1Y} is assumed for the magnetic field structure.
Since its analytical solution is available, the phase-resolved polarization can be studied in detail.
The rotating magnetic dipole moment is described by
\begin{eqnarray}
\bm{\mu}(t)=\mu(\sin\alpha\cos\Omega t \bm{e}_{x}+\sin\alpha\sin\Omega t \bm{e}_{y}+\cos\alpha\bm{e}_{z}), 
\label{eq:rotvacdipole1}
\end{eqnarray}
where $\alpha$ is the inclination angle of the magnetic dipole moment from the rotational axis of the star, $t$ is time, $\Omega$ is the rotational angular velocity of the star, and $z$-axis is aligned with the rotational axis of the star.
This generates the magnetic field described by
\begin{eqnarray}
\bm{B}(r,t)=&-&\left[ \frac{\bm{\mu}(t_{r})}{r^{3}}+\frac{\dot{\bm{\mu}}(t_{r})}{cr^{2}}+\frac{\ddot{\bm{\mu}}(t_{r})}{c^{2}r} \right] \nonumber\\
&+& \bm{e}_{r}\bm{e}_{r}\cdot\left[3\frac{\bm{\mu}(t_{r})}{r^{3}}+3\frac{\dot{\bm{\mu}}(t_{r})}{cr^{2}}+\frac{\ddot{\bm{\mu}}(t_{r})}{c^{2}r} \right],
\label{eq:rotvacdipole2}
\end{eqnarray}
where $\bm{e}_{r}$ is radial basis, $c$ is light speed, $r$ is the distance from the center of the star, and $t_{r}\equiv t-r/c$ is retarded time.  
To calculate the field lines, Eq. \ref{eq:rotvacdipole2} is integrated by the Runge-Kutta method from the stellar surface to the point of maximum cylindrical radius $R_{\rm max}$.
To ensure that the phase-resolved study adequately characterizes the polarization properties, the number of field lines in the azimuthal direction and the step length of the field lines are set to 1024 and $5\times 10^{-4} R_{\rm LC}$, respectively.
The ratio of the stellar radius to the light cylinder is set to $6.4\times 10^{-3}$, corresponding to the Crab pulsar's rotation period ($33~{\rm ms}$).

We assume outer gap as the emission region.
For the height of the emission region, the parameter $a$ is introduced, which represents the relative height of the emission region based on the last open field line (LOFL) which is the magnetic field line at the boundary between open and closed field lines.
The angle $\theta_{\rm m}$ is defined as the polar angle of the magnetic field lines measured from the magnetic axis on the stellar surface.
The polar angle on the stellar surface of the emission region is described by $\theta_{\rm m,u}=a\theta_{\rm m,pc}$, where $\theta_{\rm m,pc}$ denotes the polar angle on the stellar surface of the LOFL.
The magnetic field near the light cylinder would not be described by the current vacuum dipole magnetic field, because of the influence of the electric current.
To avoid such a theoretical uncertainty, we limit a cylindrical radius of the emission region $R_{\rm max}=0.9R_{\rm LC}$ and a radial distance $r_{\rm max}=R_{\rm LC}$, where the light cylinder radius is $R_{\rm LC}\equiv c/\Omega$. 
For the fiducial cases, the inner boundary is set to the NCS.
In the outer gap model, the inner boundary of the emission region may be located within the NCS \citep{2001ApJ...558..216H, 2004MNRAS.354.1120T}.
Since inward emission from inside the NCS significantly affects the characteristics of the light curve \citep{2008MNRAS.386..748T}, the case where the inner boundary lies inside the NCS for inward emission is also investigated (\S\ref{subsubsec:inner boundary position}).

\subsection{Emission Direction}
\label{subsec:emission direction}
It is assumed that emissions originate from particles moving at relativistic speed; thus the emission direction is highly collimated along the direction of the particle velocity.
To calculate the direction of synchrotron emission from particles, the emission direction is given by three components along the magnetic field, the gyration component perpendicular to the magnetic field, and the corotational component.
Outward emission is described by
\begin{eqnarray}
\bm{n}=(\beta_{0}\cos\theta_{\rm p})\bm{b}+(\beta_{0}\sin\theta_{\rm p})\bm{b}_{\perp}+\beta_{\rm co}\bm{e}_{\phi},
\label{eq:radvec1}
\end{eqnarray}
and inward emission is described by
\begin{eqnarray}
\bm{n}=-(\beta_{0}\cos\theta_{\rm p})\bm{b}+(\beta_{0}\sin\theta_{\rm p})\bm{b}_{\perp}+\beta_{\rm co}\bm{e}_{\phi},
\label{eq:radvec2}
\end{eqnarray}
where $\bm{b}$ is the unit vector along the magnetic field, $\bm{e}_{\phi}$ is azimuthal basis of the rotation axis, $\beta_{\rm co}\equiv R\Omega/c$ represents the corotation velocity, $\beta_{0}$ is determined by $|\bm{n}|=1$, and $\theta_{\rm p}$ is the pitch angle.
In the pulsar magnetosphere, since the particles (electron and positron pairs) which emit synchrotron emission are created by gamma-ray photons that travel their mean free path of the pair-creation process, the angle between the emission direction and the magnetic field is non-zero, which corresponds to the pitch angle $\theta_{\rm p}$.
The pitch angle $\theta_{\rm p}$ is described approximately \citep{2007ApJ...656.1044T} by
\begin{eqnarray}
\sin\theta_{\rm p}(r)=\sin\theta_{\rm p}(R_{\rm LC})\sqrt{\frac{r}{R_{\rm LC}}}.
\label{eq:pitchangle1}
\end{eqnarray}
The pitch angle value at the light cylinder, $\sin\theta_{\rm p}(R_{\rm LC})$, is treated as a model parameter with a fiducial value of $\sin\theta_{\rm p}(R_{\rm LC})=0.06$, and its dependency is discussed in \S\ref{subsec:dependence on pitch angle}.
The unit vector perpendicular to the magnetic field due to gyration $\bm{b}_{\perp}$ is described by 
\begin{eqnarray}
\bm{b}_{\perp}\equiv \pm[(\cos\delta\phi)\bm{k}+(\sin\delta\phi)\bm{k}\times\bm{b}],
\label{eq:bperp1}
\end{eqnarray}
where $\bm{k}\equiv (\bm{b}\cdot\bm{\nabla})\bm{b}/|(\bm{b}\cdot\bm{\nabla})\bm{b}|$ is the unit vector of the magnetic field line curvature, $\delta\phi$ is the gyration phase of particles, and the signs "$+$" and "$-$" represent the gyration of positron and electron, respectively. 
The phase $\delta\phi =0$ is defined when the gyration component of emission vector $\bm{b}_{\perp}$ is parallel to the curvature vector of the magnetic field line, and $\delta\phi$ increases with the direction of gyration for positrons.
The emissions are calculated by summing the contributions from all gyration phases.
Additionally, only positron is considered in the calculation, since the results are the same whether only positrons or both positrons and electrons are included, as long as emissions from all gyration phases are summed.

The light curves are calculated as a function of the viewing angle.
The viewing angle $\xi$ which is the angle of the emission vector from the rotation axis, and rotational phase $\Phi$ for each emissions are described by the following equations, 
\begin{eqnarray}
\xi=\cos^{-1}n_{z},
\label{eq:viewing_angle01}
\end{eqnarray}
\begin{eqnarray}
\Phi=-\Phi_{n}-\frac{\bm{r}\cdot\bm{n}}{R_{\rm LC}},
\label{eq:rotational_phase01}
\end{eqnarray}
where $n_{z}$ is the $z$-component of the emission vector $\bm{n}$. 
In Eq. \ref{eq:rotational_phase01}, the first term $-\Phi_{n}$ denotes the azimuthal angle of the emission vector, and the second term denotes the time delay due to the light path length.

\subsection{Stokes Parameters}
\label{subsec:stokes parameters}
Linear polarization of synchrotron emission is considered.
The observed polarization direction corresponds to the direction of the electric field vector projected onto the plane perpendicular to the line of sight.
The electric field of emission from accelerated charged particles lies in the plane parallel to the acceleration direction of the particles.
Due to gyration, the unit vector of the electric field of the emissions is described by
\begin{eqnarray}
\bm{E}_{\rm em}=(-\sin\delta\phi)\bm{k}+(\cos\delta\phi)\bm{k}\times\bm{b}.
\label{eq:accvec1}
\end{eqnarray}
The electric field projected onto the plane perpendicular to the line of sight is described by $\bm{E}_{\rm em,p}=\bm{E}_{\rm em}-(\bm{E}_{\rm em}\cdot\bm{n})\bm{n}$. 
This direction corresponds to the observed polarization direction.

To calculate the Stokes parameters, the reference for the PA in the current study is defined as the direction of the rotation axis projected onto the plane perpendicular to the line of sight $\bm{\Omega}_{\rm p}=\bm{\Omega}-(\bm{\Omega\cdot\bm{n}})\bm{n}$. 
The PA is described by 
\begin{eqnarray}
\chi^{i}=\cos^{-1}\left[\bm{E}_{\rm em,p}\cdot\frac{\bm{\Omega}_{\rm p}}{\Omega_{\rm p}}\right],
\label{eq:localpa1}
\end{eqnarray}
where $i$ means the $i$-th volume element in the emission region.
Each synchrotron photon has the PD of $\Pi_{\rm syn}=(p+1)/(p+7/3)$ where $p$ is the power law index of the particle distribution \citep{1979rpa..book.....R}. 
We apply the index $p=2$, which yields the synchrotron spectrum being consistent with the X-ray observations of the Crab pulsar.
Stokes parameters for each emission are described by the following equations,
\begin{eqnarray}
Q^{i}=\Pi_{\rm syn}I^{i}\cos2\chi^{i},
\label{eq:stokes_parameter_Q01}
\end{eqnarray}
\begin{eqnarray}
U^{i}=\Pi_{\rm syn}I^{i}\sin2\chi^{i},
\label{eq:stokes_parameter_U01}
\end{eqnarray}
where $I^{i}$ is the intensity of the emission from each volume element and gyration phase, assumed to be spatially uniform throughout the emission region for simplicity.
The total Stokes parameters observed at each viewing angle and rotational phase are calculated by adding all contributions as $I(\xi, \Phi)=\sum_{i}I^{i}$, $Q(\xi, \Phi)=\sum_{i}Q^{i}$, and $U(\xi, \Phi)=\sum_{i}U^{i}$.
The PA and PD at each viewing angle and rotational phase are calculated from the summed Stokes parameters,
\begin{eqnarray}
{\rm PA}(\xi,\Phi)=0.5\tan^{-1}\bigg[\frac{U(\xi,\Phi)}{Q(\xi,\Phi)}\bigg],
\label{eq:ppa01}
\end{eqnarray}
\begin{eqnarray}
{\rm PD}(\xi,\Phi)=\frac{\sqrt{Q^{2}(\xi,\Phi)+U^{2}(\xi,\Phi)}}{I(\xi,\Phi)}.
\label{eq:pd01}
\end{eqnarray}

The emission intensity, the PA, and the PD in the cases where the parameters are the inclination angle $\alpha$ ranging from $10^{\circ}$ to $80^{\circ}$ and the emission region height $a=0.90, 0.95$ are calculated.
In this paper, the viewing angle is basically restricted to the range of $90^{\circ}$ to $180^{\circ}$ due to the symmetry between the northern and southern hemispheres.

\subsection{Polarization Characteristics}
\label{subsec:polarization characteristics}
As seen in Fig. \ref{fig:definition} and described in \S\ref{sec:results}, a high PD and a PA variation (PA swing) at intensity peaks are expected for the following reasons. 
The synchrotron emission is hollow cone-shaped emission due to gyration of emitting particles and the local polarization directions are different for each gyration phase.
The emissions and polarization vectors are directed into a similar direction at intensity peak phases, resulting the high PD and rapid PA change at the intensity peak phases.
 
In phases where the pulsar emission is weak (i.e., the off-pulse phase), the relative contribution from the pulsar wind nebula (PWN) becomes larger, leading to increased systematic uncertainties in the polarization measurements. Therefore, in this paper, we focus on the polarization characteristics at the intensity peaks, defined by the full width at half maximum (FWHM), in order to minimize the effects of PWN contamination.

To focus the average PD and PA swing at the intensity peaks, we define the intensity peaks quantitatively as follows.
First, intensity peaks are local maxima of the light curve where the peak intensity exceeds twice the minimum intensity value (referred to as the bridge) within a single rotational period.
Second, we require that the photon counts at the peaks exceed a threshold, defined based on the convergence of PD and PA with respect to the number of magnetic field lines. 
Specifically, for a given value of $a$, we evaluate the differences in PD and PA between two cases with 512 and 1024 field lines, and adopt a threshold corresponding to typical values where the PD difference is less than $1\%$ and the PA difference is less than $4^{\circ}$ within the full width at half maximum (FWHM).
Third, if more than three intensity peaks are identified for outward emission, the two peaks with the first and second highest intensities are selected.
Fourth, these are then classified as the early peak (EP) or the late peak (LP), depending on whether they occur before or after phase 0.4 for outward emission and phase 0.7 for inward emission (Fig. \ref{fig:definition}). 
Note that the phase zero is defined as the phase when the photon emitted to the observer from the central star at which the rotation axis, magnetic axis, and the observer lie in the same plane is observed, and there is always a single intensity peak in the early ($0.0-0.4$) and late phase ($0.4-1.0$) for outward emission.

The PA swing at each intensity peak is defined as the difference between the maximum and minimum PA within the FWHM of the intensity peaks.
The PD at the intensity peak is also defined as the average PD within the FWHM. 
Additionally, the PD in the bridge region is defined as the central half of the phase interval between the EP and LP (i.e., from 1/4 to 3/4 of their phase separation of the two peaks). 
The PA swing at the bridge phase is not calculated, as the PA difference is above our criterion $4^{\circ}$ due to low photon counts.
We note that the phase resolved analysis of the polarization depends on choice of phase-bin size due to the depolarization effect.
In this study, we adopt 1/180 phase for each bin.
This width is narrower than that typically used in the observation to ensure sufficient photon statistics.
Therefore, our results will correspond to the upper limit of the observed PA swing and the PD.

Note that the polarization characteristics are almost independent of the pulsar's spin period and the strength of the magnetic field, as no energy dependency is included in the present analysis. 
Our treatment would be valid as long as the polarization characteristic of the slow-cooling regime is discussed.

\section{Results}
\label{sec:results}

\subsection{Outward Emission}
\label{subsec:outward emission}
\begin{figure}[htbp]
 \centering
 \includegraphics[width=8.5cm,clip]{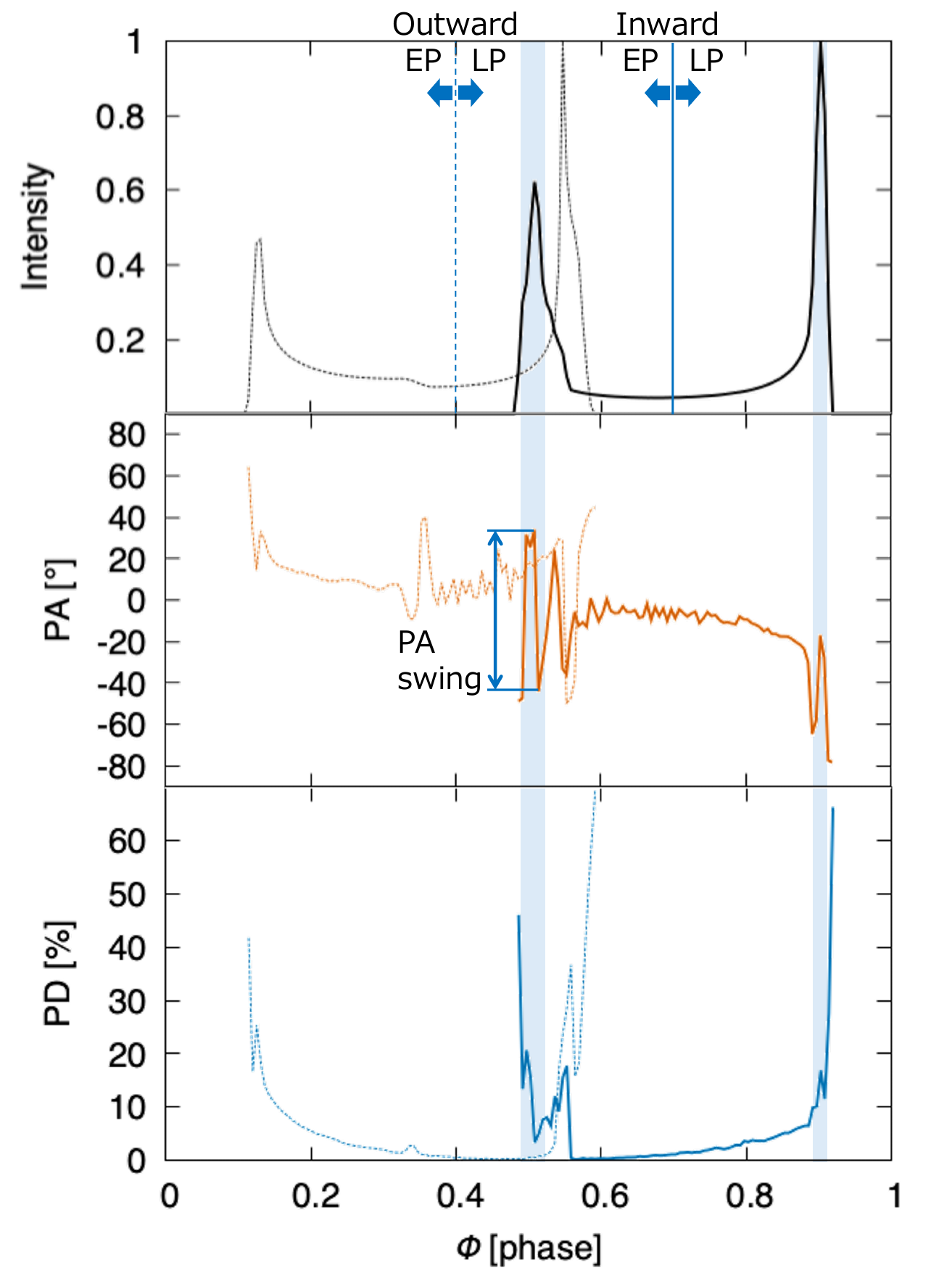}
 \caption{Intensity, PA, and PD at each rotational phase for outward and inward emission. The dotted lines show the outward emission and the solid lines show inward emission. The parameters are $\alpha=65^{\circ}$, $a=0.95$, and $\xi=100^{\circ}$. This also shows the quantitative definition of the EP, the LP, the width of the intensity peaks, and the PA swing. Phase zero occurs when a photon emitted from the star's center reaches the observer with the rotation axis, magnetic axis and observer lying in the same plane. }
 \label{fig:definition}  
\end{figure}

Fig. \ref{fig:definition} summarizes the polarization characteristics for the parameters of the inclination angle of $\alpha=65^{\circ}$, the viewing angle of $\xi=100^{\circ}$, and the emission region height of $a=0.95$ as an example of the results.
The figure shows the intensity (upper panel), the PA (middle panel), and the PD (lower panel) as a function of rotational phase for the outward emission (the dotted curve) and the inward emission (\S\ref{subsec:inward emission}) (the solid curve).
For outward emission, two intensity peaks appear and the PD tends to be higher with the PA varying significantly at the intensity peaks than the value in the bridge region between the intensity peaks (Fig. \ref{fig:definition}). 
These results are consistent with previous studies \citep{2007ApJ...656.1044T, 2017ApJ...840...73H}.

\begin{figure*}[htbp]
 \centering
 \includegraphics[width=15cm,clip]{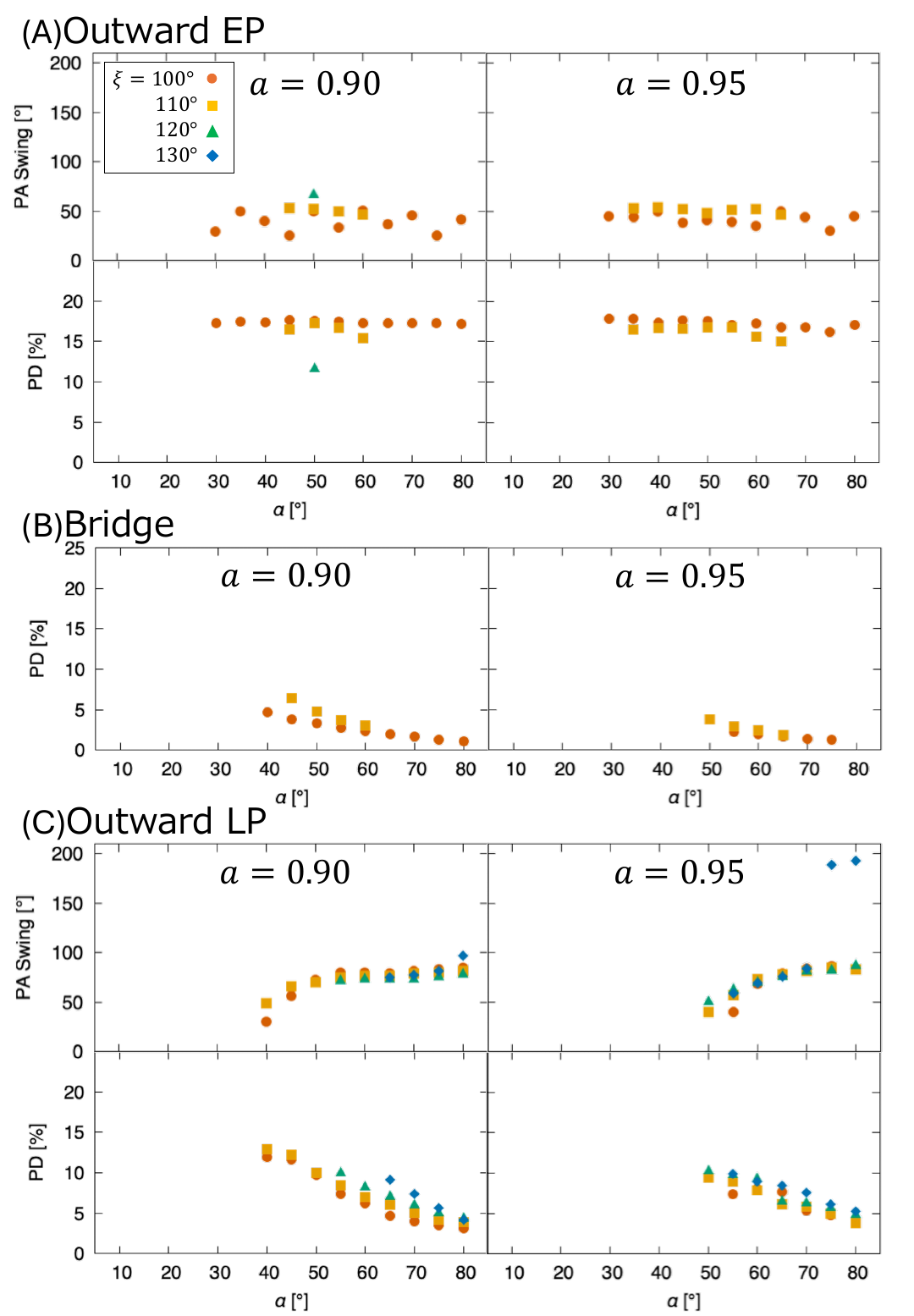}
 \caption{Parameter dependence of PA swing (top) and PD (bottom) for the EP (A), bridge (B), and LP (C) as functions of inclination angle $\alpha$ and viewing angle $\xi$ for outward emission. The emission region height is set to $a=0.90$ (left) and $a=0.95$ (right).}
 \label{fig:results_outward_dependence_alla}  
\end{figure*}

Fig. \ref{fig:results_outward_dependence_alla} shows how PA swing and the PD (the definitions are described in \S\ref{subsec:polarization characteristics}) of the outward emission depends on the inclination angle $\alpha$ and the emission region height $a$.
Panels (A), (B), and (C) correspond to the polarization of EP, bridge, and LP emission, respectively.
The figures show the results for the viewing angle of $\xi=100^{\circ}$, $110^{\circ}$, $120^{\circ}$, and $ 130^{\circ}$, above which no peak or emission can be found.
As illustrated for $\xi=120^{\circ}$ and $130^{\circ}$, if only one peak can be identified, no plot in the bridge is presented (Fig. \ref{fig:results_outward_dependence_alla} (B)).
The figures show that the PA swing typically ranges from $25^{\circ}$ to $55^{\circ}$ for the EP and from $30^{\circ}$ to $90^{\circ}$ for the LP (Fig. \ref{fig:results_outward_dependence_alla} (A) and (C)). 
Some LPs with $a=0.95$, $\alpha\ge 75^{\circ}$, and $\xi=130^{\circ}$ have large PA swing of about $180^{\circ}$.
Since their orbits of the polarization vector on the Q-U plane moves near the origin when the PD is very low, the PA is poorly defined and it changes significantly (nearly $180^\circ$).
The PD is the highest at the EP (approximately $15-20{\rm \%}$), followed by the LP ($\lesssim 15\%$), and the lowest in the bridge region ($\lesssim 5\%$).
We have examined values of $a$ ranging from $0.80$ to $1.00$. However, the dependence of the PA swing and the PD on the inclination angle is not significantly different among different values of emission height $a$; with the differences in PD and PA swing being typically less than $\sim 2\%$ and $\sim 10^{\circ}$, respectively.
Hence, in the following section, we represent the polarization characteristics only for $a=0.90$ and $0.95$.

\subsection{Inward Emission}
\label{subsec:inward emission}
\subsubsection{Intensity and Polarization Characteristics}
\label{subsubsec:intensity and polarization characteristics}
The solid lines in Fig. \ref{fig:definition} show the intensity (top panel), the PA (middle panel), and the PD (bottom panel) for the inward emission.
As illustrated in the figure, the inward emissions can also produce the two peaks per rotation \citep{2008MNRAS.386..748T}, and they generally appears within the phase range of 0.4-1.0.
The difference in the intensity peak phases at which inward and outward emissions is explained by effects of light path length and corotation velocity. 
The direction of inward emission is opposite to that of outward emission (Eq. \ref{eq:radvec2}).
This results in the outward and inward emission regions being located forward and backward of the star, respectively, respect to the observer.
Also, the direction of the corotation velocity of the inward emission is opposite to that of the outward emission.
Consequently, the difference in the intensity peak phases of the outward and inward emissions arises.
We found that inward emission can contribute significantly to the phase interval usually identified as the off-pulse phase in models considering only outward emission (Fig. \ref{fig:definition}).
The minimum of the observed phase-resolved emission intensity, corresponding to the emission in the off-pulse phase, is assumed to arise solely from the PWN as a simple model.
It should be noted, however, that the off-pulse component may still contain contributions from the pulsar itself, which can introduce systematic errors in simple subtraction.

\begin{figure*}[htbp]
 \centering
 \includegraphics[width=13.1cm,clip]{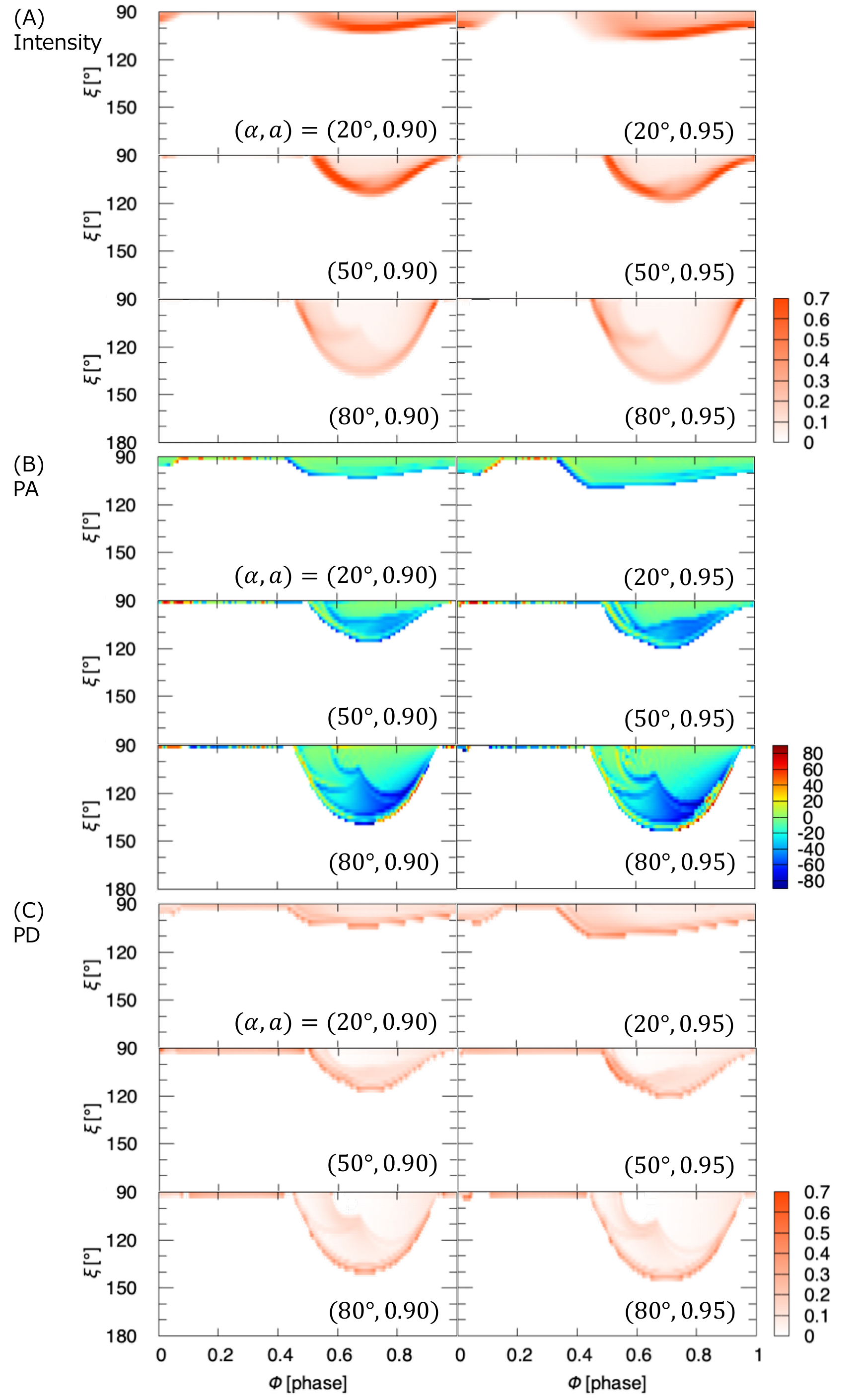}
 \caption{Emission intensity (panel A), PA (panel B), and PD (panel C) as a function of viewing angle and rotational phase for inward emission. The viewing angle is limited from $90^{\circ}$ to $180^{\circ}$ due to north-south symmetry. To emphasize the intensity peaks, the maximum value of the color bar is set to 0.7, although the actual maximum intensity is 1. The color bar of PA shows PA value in unit of degree. The left and right columns show $a=0.90$ and $0.95$, respectively. The top, middle, and bottom rows of each panel show $\alpha=20^{\circ}$, $50^{\circ}$, and $80^{\circ}$, respectively.}
 \label{fig:results_inward_skymap}  
\end{figure*}

At the intensity peaks, the PA exhibits significant variation and the PD tends to be high in the middle and bottom panels of Fig. \ref{fig:definition}, respectively.
Fig. \ref{fig:results_inward_skymap} shows the sky map of the polarization characteristics of the inward emissions; panels (A), (B), and (C) correspond to the intensity, PA, and PD, respectively.
The parameter ranges of the inclination angle $\alpha$, the viewing angle $\xi$, and the height of the emission region $a$ are the same as in the outward emission.
As illustrated in Fig. \ref{fig:results_inward_skymap} (A), the phase separation between two intensity peaks increases with larger inclination angle and lower emission height (i.e., larger values of $a$), and decreases as the viewing angle increases from $90^{\circ}$, the same as in the behavior confirmed in the outward emission \citep{2007ApJ...656.1044T, 2010ApJ...715.1270B, 2017ApJ...840...73H}.
The characteristics which the PA exhibits significant variation and the PD tends to be high at the intensity peaks (Fig. \ref{fig:definition}) are also seen in most parameter sets (Figs. \ref{fig:results_inward_skymap} (B) and (C)).

\begin{figure*}[htbp]
 \centering
 \includegraphics[width=15cm,clip]{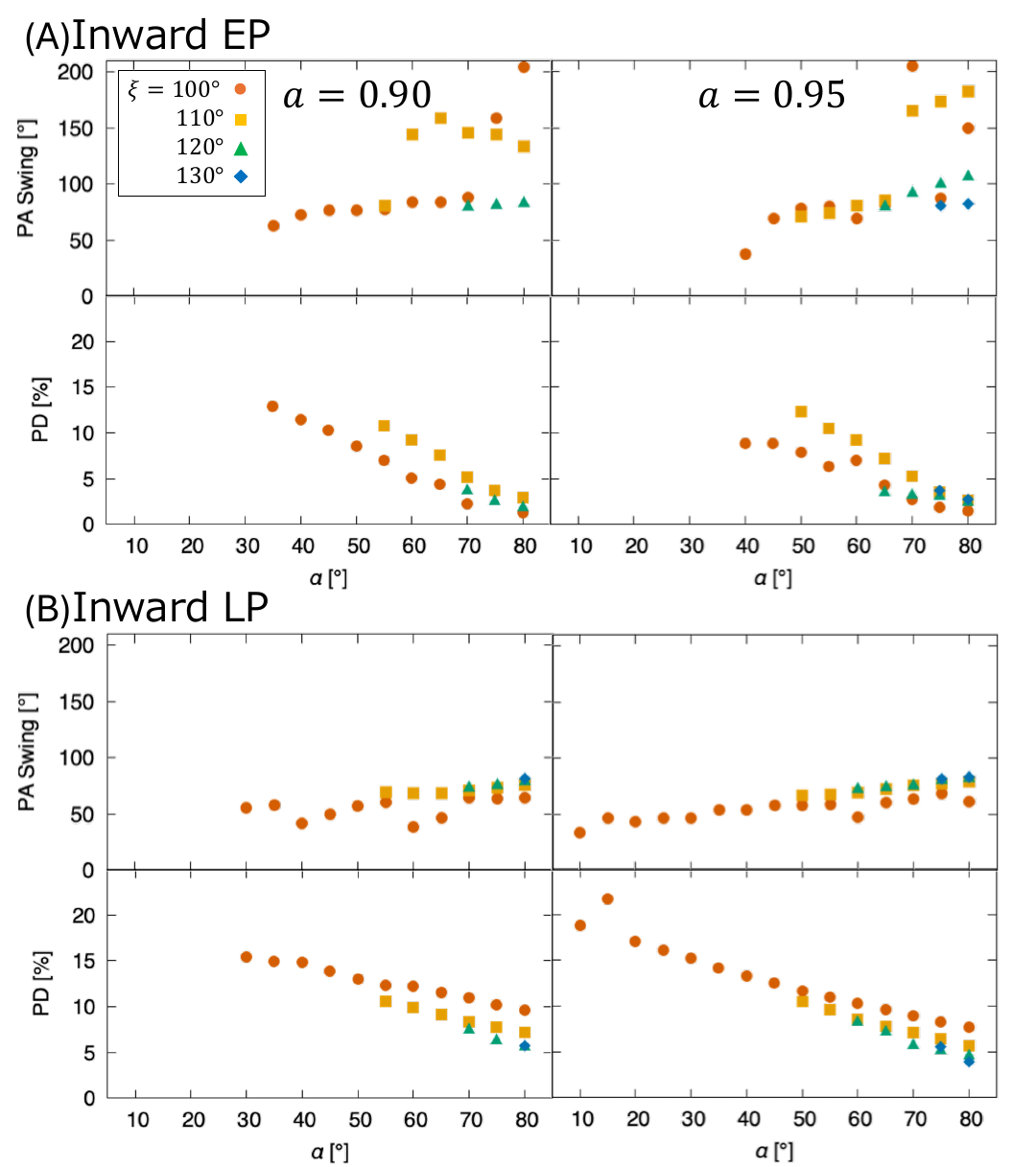}
 \caption{Parameter dependence of PA swing (top) and PD (bottom) for the EP (panel A) and the LP (panel B), as functions of inclination angle $\alpha$ and viewing angle $\xi$ for inward emission. The height parameter of the emission region is set to $a=0.90$ (left) and $a=0.95$ (right).}
 \label{fig:results_inward_dependence_alla}  
\end{figure*}

As in Fig. \ref{fig:results_outward_dependence_alla} of the outward emission, Fig. \ref{fig:results_inward_dependence_alla} shows how the polarization characteristics of EP (panel A) and LP (panel B) of the inward emission depend on the geometry ($\alpha$, $\xi$, and $a$).
From Fig. \ref{fig:results_inward_dependence_alla}, the PA swing is $\gtrsim 50^{\circ}$ for the EP (panel A), and ranges from $20^{\circ}$ to $90^{\circ}$ for the LP (panel B). 
The PD decreases as the inclination angle increases.
The PD for EP ($\lesssim15{\rm \%}$) (panel A) is relatively lower than that for LP ($\lesssim 20{\rm \%}$) (panel B).
The EPs with $a=0.90$ and $0.95$, $\alpha\ge 60^{\circ}$, and $\xi=100^{\circ}-110^{\circ}$ have a large PA swing of $>130^{\circ}$.
Similarly to the outward emission (Fig. \ref{fig:results_outward_dependence_alla}), since their orbits of the polarization vector on the Q-U plane moves near the origin, the PA swing changes significantly.

\subsubsection{Inner Boundary Position of Emission Region}
\label{subsubsec:inner boundary position}
\begin{figure}[htbp]
 \centering
 \includegraphics[width=8.5cm,clip]{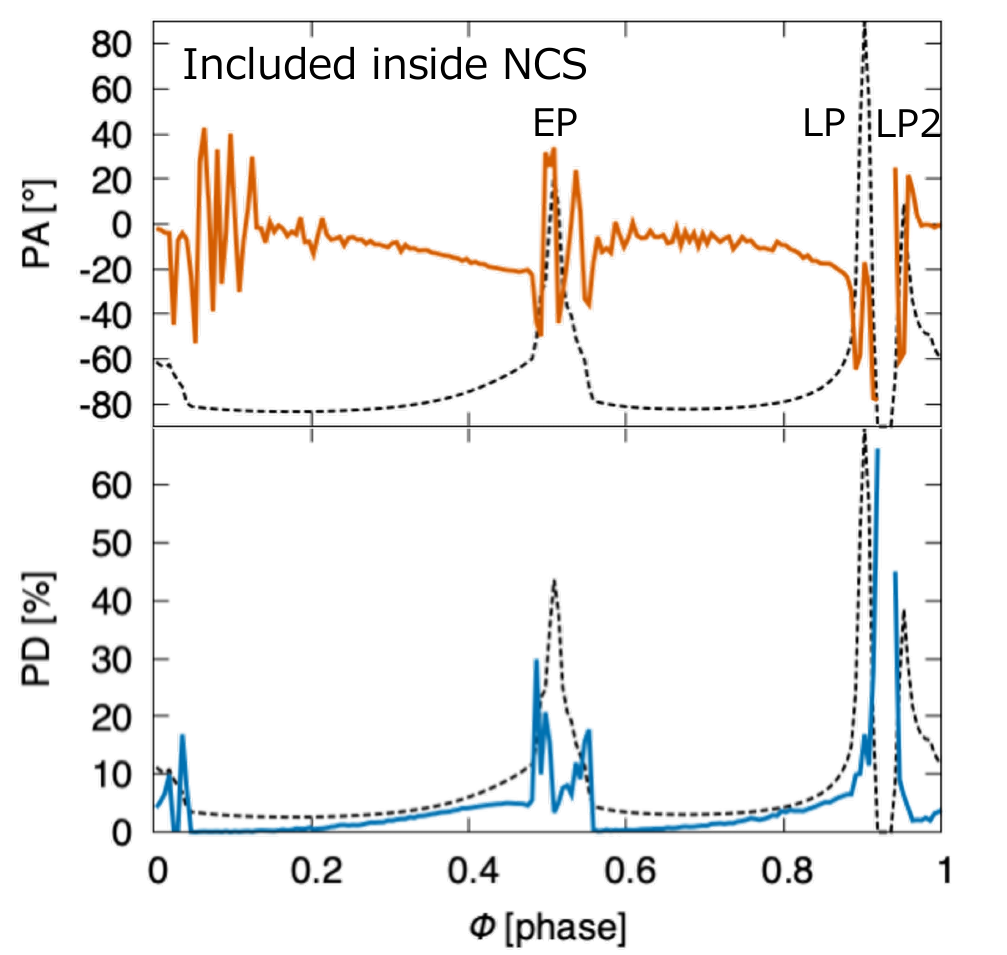}
 \caption{PA (top) and PD (bottom) for inward emission. Black dotted line shows the intensity. This shows the case where emission region includes the inside NCS (LP2). The parameters are $\alpha=65^{\circ}$, $a=0.95$, and $\xi=100^{\circ}$.}
 \label{fig:results10_inner_boundary}  
\end{figure}

In this study, we adopt the position of the inner boundary as a model parameter, since it is determined by the electrodynamics of the global magnetosphere \citep{2003ApJ...591..334H, 2008MNRAS.386..748T}.
For Figs. \ref{fig:definition} - \ref{fig:results_inward_dependence_alla}, we assumed that the inner boundary is located at the NCS.
To illustrate the dependency of the polarization characteristics on the boundary position, Fig. \ref{fig:results10_inner_boundary} shows the result for the inner boundary position of stellar surface; the parameters are for $\alpha=65^\circ$, $a=0.95$, and $\xi=100^\circ$.
Inward emission from inside NCS typically contributes to the intensity in the phase range 0.0-0.5 and forms an intensity peak near 1.0 phase (Fig. \ref{fig:results10_inner_boundary}), which does not overlap with the outward intensity peaks.
In Fig. \ref{fig:results10_inner_boundary}, this peak created by the inward emission is indicated by the label "LP2".
Since the emission region responsible for forming LP2 (inside the NCS) is located closer to the star than those producing LP and EP (outside the NCS), the emitting particles that form LP2 have smaller pitch angles (Eq. \ref{eq:pitchangle1}).
As a result, photons with different polarization angles overlap in a narrow phase, leading lower PD compared to inward EP and LP, as Fig. \ref{fig:results10_inner_boundary} shows.

\begin{figure*}[htbp]
 \centering
 \includegraphics[width=15cm,clip]{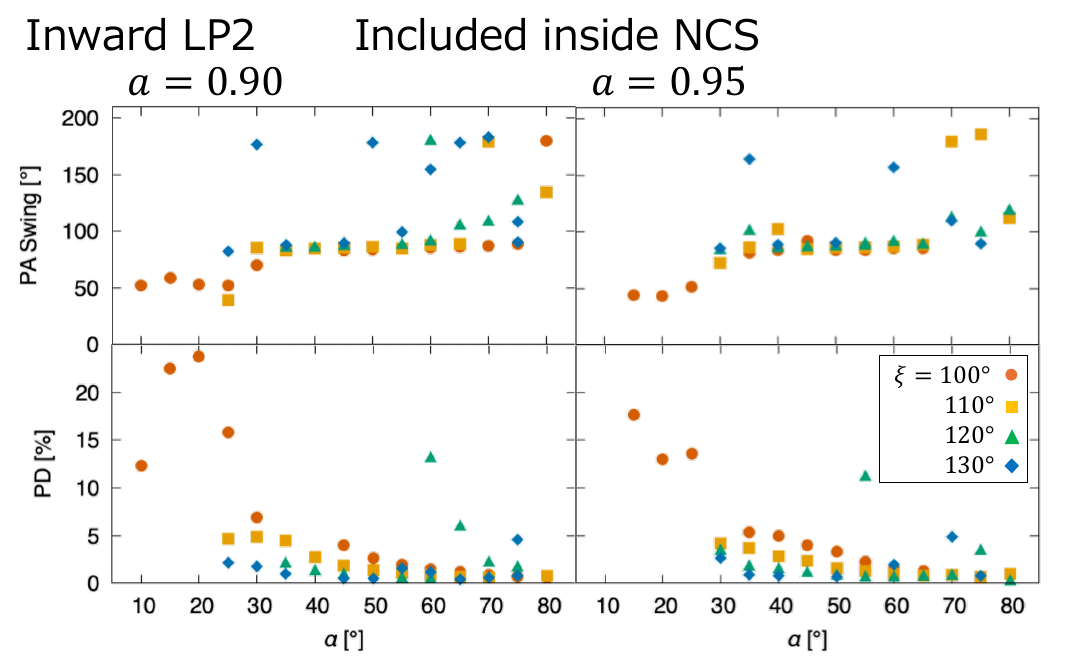}
 \caption{The vertical axes, the horizontal axes, and the symbols are same with Fig. \ref{fig:results_inward_dependence_alla}, however this figure is for LP2.}
 \label{fig:inward_dependence_innerb}  
\end{figure*}

Fig. \ref{fig:inward_dependence_innerb} shows how the PA swing and the PD for the LP2 depend on the geometry parameters $\alpha, \xi$, and $a$.
In the figure, although the inward emission beyond NCS may contribute to LP2, its effect on the PD and the PA swing is negligible for the inclination angle of $\alpha>25^{\circ}$.
As Fig. \ref{fig:inward_dependence_innerb} shows, the PA swing is larger ($\sim70^{\circ}-100^{\circ}$) and the PD is lower ($\lesssim5\%$) for the LP2.
In particular, the PA swing over the LP2 can reach $\sim 180^{\circ}$ such as $\alpha=30^{\circ}$, $35^{\circ}$, $50^{\circ}$, and $\ge60^{\circ}$ (Fig. \ref{fig:inward_dependence_innerb}).
When the PD is very low as $\lesssim3\%$, the PA is poorly defined.
As a result, even a slight change in the relative contributions of the polarization components can cause a large change in the PA, producing a significant swing (nearly $180^\circ$).
There are LP2 with the inclination angle $\alpha \lesssim 25^{\circ}$ which have the PD of $\gtrsim10\%$.
These correspond to cases where the contribution of emission from NCS outside is significant.

\subsubsection{Dependence on Pitch Angle}
\label{subsec:dependence on pitch angle}
\begin{figure*}[htbp]
 \centering
 \includegraphics[width=15.9cm,clip]{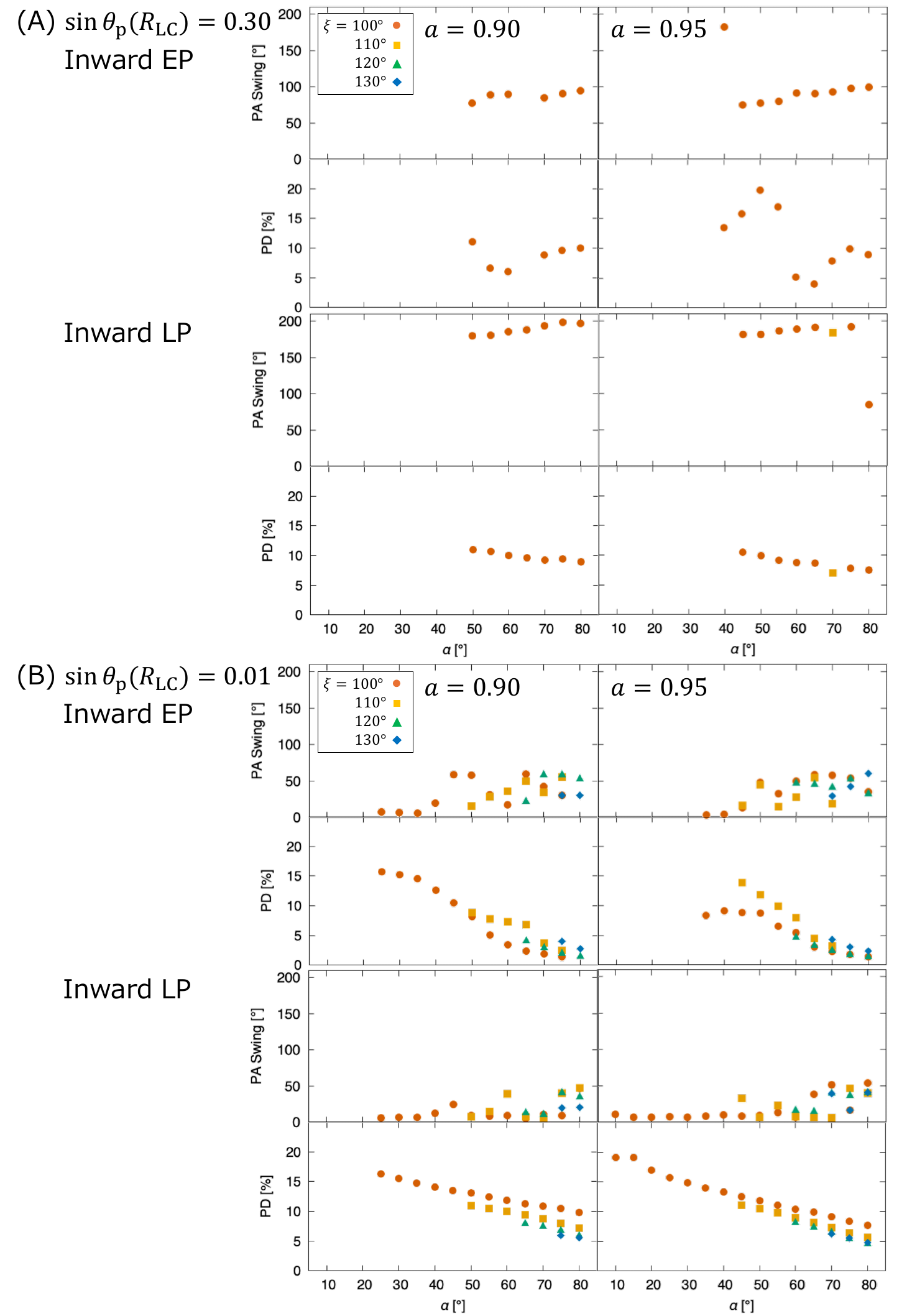}
 \caption{The vertical axes, the horizontal axes, and the symbols are same with Fig. \ref{fig:results_inward_dependence_alla}, however these show for pitch angles of $\sin\theta_{\rm p}(R_{\rm LC})=0.30$ (panel A) and $0.01$ (panel B).}
 \label{fig:pitch0.3_in_dependence}  
\end{figure*}

Fig. \ref{fig:pitch0.3_in_dependence} shows how the PA swing and the PD depend on the pitch angle; $\sin\theta_{\rm p}(R_{\rm LC})=0.30$ for panel A and $0.01$ for panel B.
Compared to the case of $\sin\theta_{\rm p}(R_{\rm LC})=0.06$ in Fig. \ref{fig:results_inward_dependence_alla}, it is found that the larger pitch angle leads to the larger PA swing and the slightly higher or comparable PD.
Specifically, the PA swing is $\sim0^{\circ}-50^{\circ}$ at $\sin\theta_{\rm p}(R_{\rm LC})=0.01$ (Fig. \ref{fig:pitch0.3_in_dependence} (B)), $\sim50^{\circ}-100^{\circ}$ at $\sin\theta_{\rm p}(R_{\rm LC})=0.06$ (Fig. \ref{fig:results_inward_dependence_alla} (B)), and $\sim150^{\circ}-200^{\circ}$ at $\sin\theta_{\rm p}(R_{\rm LC})=0.30$ (Fig. \ref{fig:pitch0.3_in_dependence} (A)) for the LP.
This tendency of increase in the PA swing with larger pitch angle can be understood as follows.
For a given rotation phase width, a larger pitch angle allows a synchrotron emission from broader region to contribute, results in wider variation of the polarization angle.
The PD does not vary significantly with the pitch angle and it remains $\lesssim 20\%$ for $\sin\theta_{\rm p}(R_{\rm LC})=0.01$, $0.06$, and $0.30$.

\subsection{The Overlapping Peak}
\label{subsec:the_case_outward_and_inward_emission_overlap}
In general, both outward and inward emissions are expected to be present. 
At the phase of an intensity peak associated with either component, the polarization properties are typically dominated by that component even when both contributions are summed. 
However, the phases of the intensity peaks of the two components may overlap, in which case both contributions must be taken into account. 
In such cases, an X-ray intensity peak coinciding with the gamma-ray peak—generally dominated by outward emission—may be interpreted as being outward-dominated based on intensity alone, whereas polarization properties may still retain information on the inward emission component.

To investigate this effect, we examine cases in which the intensity peaks of the two components overlap and compare them with the case of outward emission alone.
For simplicity, we assume that the volume emissivities of the outward and inward emissions are identical.
Overlap is expected to occur between the outward LP and inward EP components.
We define overlapping peaks as those in which the phase of the outward LP lies within the FWHM of the inward EP, or vice versa.

\begin{figure*}[htbp]
 \centering
 \includegraphics[width=16.5cm,clip]{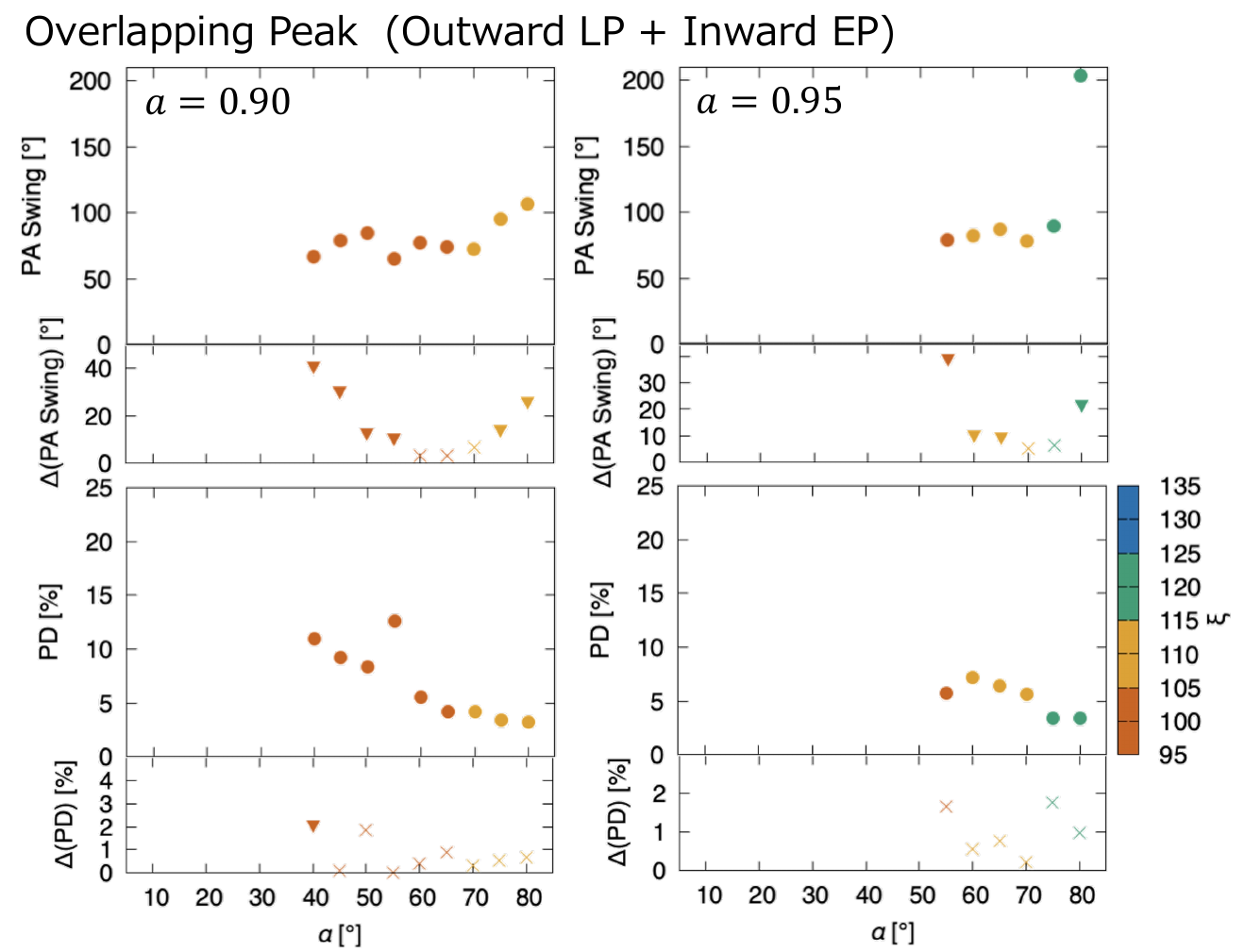}
 \caption{Parameter dependence of the PA swing (upper panels) and the PD (lower panels) for overlapping peak as functions of inclination angle $\alpha$.
 The viewing angle is plotted within the range $95^\circ \le\xi \le 135^\circ$. 
 The bottom panels show the differences between the overlapping peak and the outward LP. Cross points indicate cases where the differences of PA swing and PD are $8^{\circ}$ and $2\%$ or less, respectively. 
 Note, the PA causing the abnormal PA swing is replaced to the PA at the phase with the larger emission intensity in the adjacent phase, in the three cases where ($a, \alpha, \xi$)=($0.90, 50^{\circ}, 94^{\circ}$), ($0.90, 80^{\circ}, 110^{\circ}$), and ($0.95, 80^{\circ}, 110^{\circ}$) since they deviate significantly from the overall trend, and considering the model's uncertainty, they are not strongly expected to be observed in reality.}
 \label{fig:results_overlap}  
\end{figure*}

The top panel of Fig. \ref{fig:results_overlap} shows the parameter dependence of the PD and the PA swing for the overlapping peaks, while the bottom panel presents the differences relative to the case of outward emission alone. 
The plotted points correspond to viewing angles that satisfy the overlap condition and maximize the PA swing difference between the overlapping and outward-only cases for each inclination angle.
We find that the PA swing increases to $\gtrsim 50^{\circ}$ in the overlapping case, while the PD remains largely unchanged. 
This behavior can be attributed to the difference in PA swing between the outward LP ($\gtrsim 30^{\circ}$; Fig.~\ref{fig:results_outward_dependence_alla} (C)) and the inward EP ($\gtrsim 50^{\circ}$; Fig.~\ref{fig:results_inward_dependence_alla} (A)), together with their similar PD values of $\lesssim 15\%$.
The inward emission from regions inside NCS does not contribute significantly at these overlapping phases.

\begin{table*}[htb]
 \caption{Expected parameters in each PD and PA swing ($\xi> 90^{\circ}$, emission region is outside NCS)}
  \centering
   \begin{tabular}{ccc|cc|cc|cc} \hline \hline
        & & & \multicolumn{2}{c|}{PA swing($^{\circ}$)$=0-50$} & \multicolumn{2}{c|}{$50-100$} & \multicolumn{2}{c}{$100-210$} \\ 
     PD($\%$) & Component & $\sin\theta_{\rm p}(R_{\rm LC})$ & $\xi$($^{\circ}$) & $\alpha$($^{\circ}$) & $\xi$($^{\circ}$) & $\alpha$($^{\circ}$) & $\xi$($^{\circ}$) & $\alpha$($^{\circ}$) \\ \hline
     \multirow{13}{*}{$0-5$} & \multirow{3}{*}{Outward EP} & 0.01 & -- & -- & -- & -- & -- & -- \\
           &  & 0.06  & -- & -- & -- & -- & -- & -- \\
           &  & 0.30  & -- & -- & -- & -- & -- & -- \\ \cline{3-9}
           & \multirow{3}{*}{Outward LP} & 0.01 & 100-130 & 65-80 & -- & -- & -- & -- \\
           &  & 0.06 & -- & -- & 100-130 & 50-80 & -- & -- \\
           &  & 0.30 & -- & -- & 100-110 & 70-80 & -- & -- \\ \cline{3-9} 
           & \multirow{3}{*}{Inward EP}  & 0.01 & 100-130 & 55-80 & 100-130 & 60-80 & --  & -- \\
           &  & 0.06 & -- & -- & 100-130 & 60-80 & 100-110 & 60-80 \\
           &  & 0.30 & -- & -- & 100 & 60-65 & -- & -- \\ \cline{3-9}
           & \multirow{3}{*}{Inward LP}  & 0.01 & -- & -- & -- & -- & -- & -- \\ 
           &  & 0.06 & -- & -- & -- & -- & -- & -- \\
           &  & 0.30 & -- & -- & -- & -- & -- & -- \\ \cline{3-9}
           & Outward LP + Inward EP & 0.06 & -- & -- & 115-125 & 75-80 & 115-125 & 80 \\ \hline
    \multirow{13}{*}{$5-10$} & \multirow{3}{*}{Outward EP} & 0.01 & -- & -- & -- & -- & -- & -- \\
           &  & 0.06  & -- & -- & -- & -- & -- & -- \\
           &  & 0.30  & -- & -- & -- & -- & -- & -- \\ \cline{3-9}
           & \multirow{3}{*}{Outward LP} & 0.01 & 100-130 & 50-80 & -- & -- & -- & -- \\
           &  & 0.06 & -- & -- & 100-130 & 50-80 & 130 & 75-80 \\
           &  & 0.30 & -- & -- & 100-110 & 60-80 & -- & -- \\ \cline{3-9} 
           & \multirow{3}{*}{Inward EP}  & 0.01 & 100-120 & 35-65 & 100-120 & 45-65 & --  & -- \\
           &  & 0.06 & -- & -- & 100-120 & 40-70 & -- & -- \\
           &  & 0.30 & -- & -- & 100 & 55-80 & -- & -- \\ \cline{3-9}
           & \multirow{3}{*}{Inward LP}  & 0.01 & 100-130 & 60-80 & -- & -- & -- & -- \\ 
           &  & 0.06 & -- & -- & 100-130 & 55-80 & -- & -- \\
           &  & 0.30 & -- & -- & 100 & 80 & 100-110 & 45-80 \\ \cline{3-9}
           & Outward LP + Inward EP & 0.06 & -- & -- & 95-115 & 45-75 & -- & -- \\ \hline
     \multirow{13}{*}{$10-15$} & \multirow{3}{*}{Outward EP} & 0.01 & 130 & 60 & -- & -- & -- & -- \\
           &  & 0.06  & -- & -- & 120 & 50 & -- & -- \\
           &  & 0.30  & -- & -- & -- & -- & -- & -- \\ \cline{3-9}
           & \multirow{3}{*}{Outward LP} & 0.01 & 100-130 & 35-65 & -- & -- & -- & -- \\
           &  & 0.06 & 100-120 & 40-50 & 100-130 & 40-50 & -- & -- \\
           &  & 0.30 & -- & -- & 100-110 & 50-80 & -- & -- \\ \cline{3-9} 
           & \multirow{3}{*}{Inward EP}  & 0.01 & 100-110 & 25-55 & -- & -- & --  & -- \\
           &  & 0.06 & -- & -- & 100-110 & 35-60 & -- & -- \\
           &  & 0.30 & -- & -- & 100 & 40-80 & 100 & 40 \\ \cline{3-9}
           & \multirow{3}{*}{Inward LP}  & 0.01 & 100-110 & 30-80 & -- & -- & -- & -- \\ 
           &  & 0.06 & -- & -- & 100-110 & 30-80 & -- & -- \\
           &  & 0.30 & -- & -- & -- & -- & 100 & 45-80 \\ \cline{3-9}
           & Outward LP + Inward EP & 0.06 & -- & -- & 95-105 & 40-55 & -- & -- \\ \hline
     \multirow{13}{*}{$15-20$} & \multirow{3}{*}{Outward EP} & 0.01 & 100-110 & 10-80 & -- & -- & -- & -- \\
           &  & 0.06  & 100-110 & 30-80 & 100-110 & 30-80 & -- & -- \\
           &  & 0.30  & -- & -- & -- & -- & -- & -- \\ \cline{3-9}
           & \multirow{3}{*}{Outward LP} & 0.01 & -- & -- & -- & -- & -- & -- \\
           &  & 0.06 & -- & -- & -- & -- & -- & -- \\
           &  & 0.30 & -- & -- & 100-110 & 55-65 & -- & -- \\ \cline{3-9} 
           & \multirow{3}{*}{Inward EP}  & 0.01 & 100 & 25-35 & -- & -- & --  & -- \\
           &  & 0.06 & -- & -- & -- & -- & -- & -- \\
           &  & 0.30 & -- & -- & 100 & 45-55 & -- & -- \\ \cline{3-9}
           & \multirow{3}{*}{Inward LP}  & 0.01 & 100 & 10-30 & -- & -- & -- & -- \\ 
           &  & 0.06 & 100 & 10-30 & -- & -- & -- & -- \\
           &  & 0.30 & -- & -- & -- & -- & -- & -- \\ \cline{3-9}
           & Outward LP + Inward EP & 0.06 & 95-105 & 45 & -- & -- & -- & -- \\ \hline
   \end{tabular}
  \label{tab:table01}
\end{table*}

\section{Discussion}
\label{sec:discussion}
We have carried out a detail investigation for polarization properties as result of the synchrotron emission from the outer gap.
The light curve and polarization properties are calculated by taking into account for both the inward and outward emissions.

Table \ref{tab:table01} summarizes the results for the PD and the PA swing investigated in the previous sections for the case where the inner boundary of the emission region locates at the NCS.
It is found that the polarization of inward emission has the PD of $\lesssim 20\%$ near the EPs, accompanied by the PA swing of over $50^{\circ}$ for the pitch angle $\sin\theta_{\rm p}(R_{\rm LC})=0.06$ (Table. \ref{tab:table01}).
It is also found that the PD for the overlapping peak of the outward LP and the inward EP is not significantly affected, and the PA swing is $\gtrsim 50^{\circ}$ since the PA swing of the inward EP tends to be larger (typically $\gtrsim 50^{\circ}$) than that of the outward LP (typically $\gtrsim 30^{\circ}$) for the pitch angle $\sin\theta_{\rm p}(R_{\rm LC})=0.06$ (Table \ref{tab:table01}).

As discussed in \S\ref{subsec:dependence on pitch angle} and seen in Table \ref{tab:table01}, a larger pitch angle leads to a larger PA swing.
Within the frame work of the outer gap, the pitch angle of the secondary pairs is associated to the mean-free path of the gamma-rays, and it increases as the mean-free path increases.
As a pulsar spun down, the mean-free path increases as the soft-photon field outer magnetosphere decreases.
Therefore, our model predicts that the pulsed emission from pulsars with lower spin-down luminosity would exhibit a larger PA swing.

Outside the outer gap model, the current sheet model is also considered a pulsed emission region.
\citet{2016MNRAS.457.2401C, 2016MNRAS.463L..89C} have calculated the polarization properties based on the particle simulation results which the emission region is mainly the current sheet, showing that a PD dip accompany the intensity peaks.
These intensity peaks occur when the line of sight crosses the current sheet where magnetic polarity changes, resulting in PD decreases due to caustics of various polarized emissions \citep{2013MNRAS.434.2636P, 2016MNRAS.463L..89C}.
A key difference is that PD increases at intensity peaks in the outer gap model for both outward and inward emissions (Fig. \ref{fig:definition}), while it decreases in the current sheet model.
This difference is attributed to the relatively aligned magnetic field in the outer gap, compared to the polarity flips in the current sheet.

Although the current treatment assumes a vacuum rotating dipole field, which may not represent the magnetic field structure near the light cylinder, this approach and the comparison between the model results and the observation data would provide useful information for inferring the actual magnetic field configuration in the pulsar magnetosphere.
It has been shown (e.g., \citet{2026A&A...706A.215P}, \citet{2010ApJ...715.1270B}, \citet{2010ApJ...715.1282B}) that the magnetic field line structure in a force-free magnetosphere is deformed in the direction opposite to the rotation, compared to the vacuum field configuration, primarily in the vicinity of and beyond the light cylinder. Consequently, the EP is expected to be more strongly affected by the differences between the vacuum and force-free magnetospheres than the LP, because its emission region is located closer to the light cylinder. In the force-free case, the magnetic field lines in the EP emission region become bent in the same direction as those in the LP emission region. As a result, the polarization characteristics of the EP are expected to approach those of the LP, whereas the differences in the LP polarization characteristics between the vacuum and force-free magnetospheres are expected to be less pronounced. 

IXPE has measured the polarization characteristics from four pulsars, the Crab pulsar \citep{2024ApJ...973..172W}, PSR B1509-58 \citep{2023ApJ...957...23R}, PSR B0540-69 \citep{2024ApJ...962...92X}, and PSR J0205+6449 \citep{2025A&A...699A..33B}.
They are discussed based on our results.
We note that the adopted phase-bin width of 1/180 is narrower than that typically used in observations, and the results correspond to the upper limit of the observed PA swing and the PD.

X-ray polarization observations of the Crab pulsar by IXPE has been reported by \citet{2023NatAs...7..602B} and \citet{2024ApJ...973..172W}.
According to X-ray polarization data \citep{2024ApJ...973..172W}, the PA increases (by approximately $40^{\circ}$), and the PD reaches about $15\%$ at the main pulse.
At the interpulse, the PA swing is around $20^{\circ}$, and the PD is $5-10\%$.
The observed PD and PA swing at the main peak in X-ray \citep{2024ApJ...973..172W} of the Crab pulsar supports that the main contribution is the outward emission (Fig. \ref{fig:definition}). 
The observed PA swing of $\sim20^{\circ}$ at the interpulse also suggests that outward emission dominates in the Crab pulsar's X-ray band, considering that inward emission typically produces PA swing at the LP approach to $\gtrsim50^{\circ}$.
If PA swing increases with improved phase resolution by future observations with higher phase resolution, the contribution of inward emission could be detected.

\begin{figure*}[htbp]
 \centering
 \includegraphics[scale=0.5,clip]{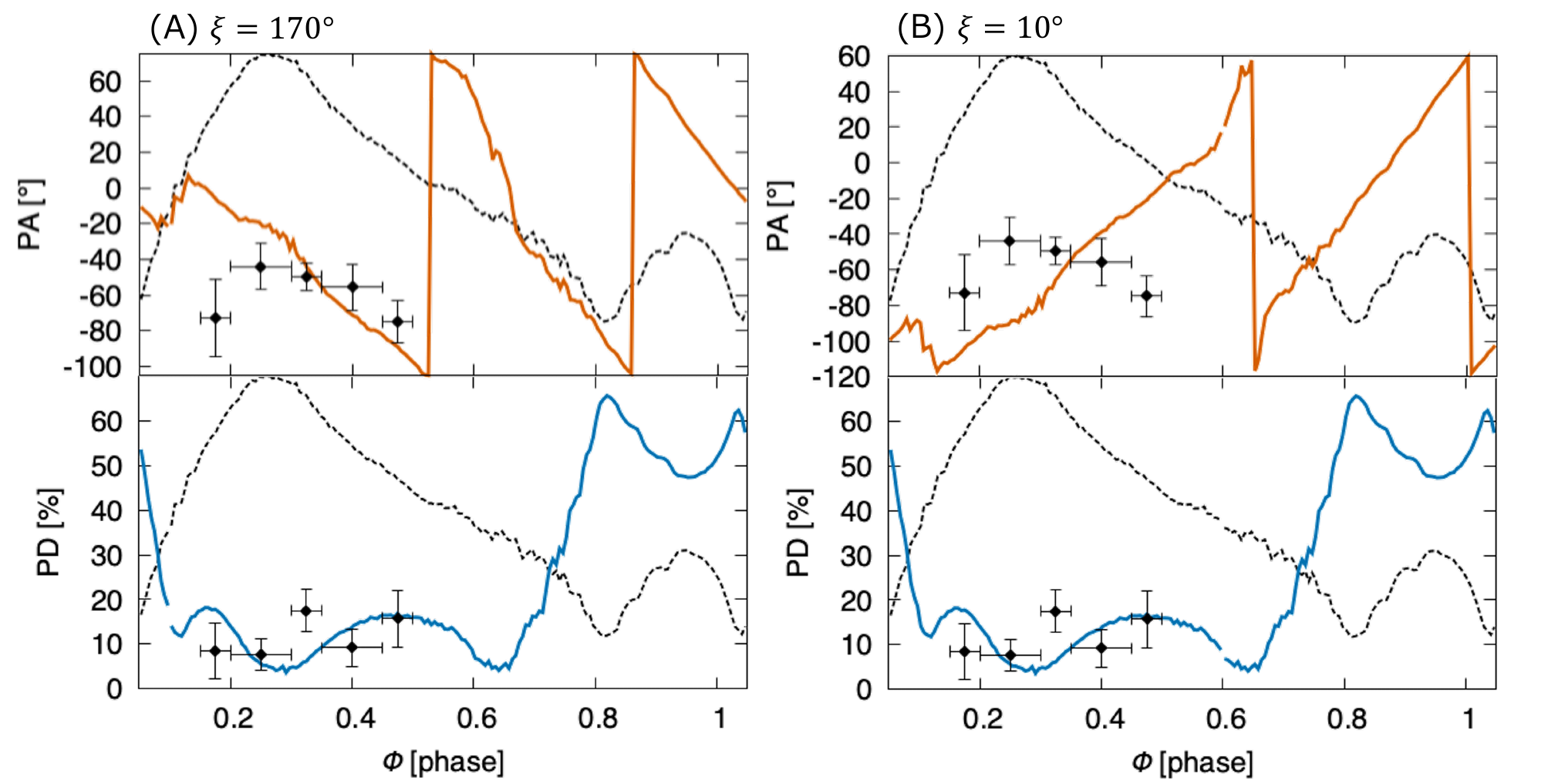}
 \caption{The red and blue lines show the calculated PA and PD at a phase resolution of 1/180. The black dashed line shows the calculated intensity. The black points represent the observed PA and PD data for PSR B1509-58 \citep{2023ApJ...957...23R}. Panels A and B show the cases with viewing angles of $\xi = 170^{\circ}$ and $\xi = 10^{\circ}$, respectively. The other parameters are $\alpha = 20^{\circ}$, $a = 0.95$, and $\sin\theta_{\rm p}(R_{\rm LC})=0.60$.}
 \label{fig:b1509}  
\end{figure*}

\citet{2023ApJ...957...23R} reported X-ray polarization measurements of PSR B1509-58.
Given its unusually soft gamma-ray spectrum \citep{1999A&A...351..119K, 2023ApJ...958..191S}, which has been interpreted in terms of the polar cap model \citep{1997ApJ...476..246H}, they fitted a rotating vector model to the radio and X-ray PA data and obtained results consistent with the observations.
An alternative interpretation based on inward emission in the outer gap has also been proposed \citep{2013ApJ...764...51W}.
Using parameters similar to those adopted in that model ($\alpha = 20^{\circ}$, $a = 0.95$, and $\sin\theta_{\rm p}(R_{\rm LC}) = 0.60$) and condition that emission occurs only along magnetic field lines where the NCS is located inside the light cylinder, we computed the expected polarization properties of inward emission for both $\xi=10^\circ$ and $170^\circ$, together with the inclusion of emission from inside the NCS in Fig. \ref{fig:b1509}.
Here, we restrict our analysis to the data within the FWHM of both the observed and modeled intensity peaks, where the effects of possible PWN contamination and limited photon statistics are expected to be less significant. 
Since the photon count at the intensity peak fell below the threshold (\S \ref{subsec:polarization characteristics}) at the same resolution used previously (Fig. \ref{fig:results_outward_dependence_alla}, \ref{fig:results_inward_dependence_alla}, and \ref{fig:inward_dependence_innerb}-\ref{fig:results_overlap}), we performed the calculation by quadrupling the number of magnetic field lines (to 4096) and making the step length of the field lines five times finer ($10^{-4}~R_{\rm LC}$).
The model phase was shifted by 0.1 to align the X-ray intensity peaks shown in Fig. \ref{fig:b1509}, since the phase of the radio pulse does not necessarily coincide with the model reference phase ($\Phi=0$). 
The results, shown in Fig. \ref{fig:b1509}, demonstrate that the model with $\xi=170^\circ$ gives a substantially better description of the observed X-ray polarization properties than the model with $\xi=10^{\circ}$, although neither model provides a statistically acceptable fit.
The constant PA offset is a free parameter, and its values are $-46^{\circ}$ and $-64^{\circ}$ for $\xi=170^\circ$ and $10^\circ$, respectively. The resulting values of $\chi^2/{\rm d.o.f.}$ are approximately 3 and 5 for $\xi=170^\circ$ and $10^\circ$, respectively.
These results suggest that, although the intensity profile alone cannot distinguish between viewing angles of $\xi$ and $180^\circ-\xi$, the X-ray polarization properties favor $\xi\sim170^\circ$ over the previously assumed $\sim10^\circ$.
Moreover, the phase dependence of the PA differs between the models: in the polar cap model of \citet{2023ApJ...957...23R}, the X-ray PA decreases with phase around the pulse peak, whereas it increases in the outer-gap inward-emission scenario. Further X-ray polarization observations with improved statistical precision will be required to distinguish between these models, and between the possible viewing angles within the outer gap model.

\begin{figure}[htbp]
 \centering
 \includegraphics[scale=0.5,clip]{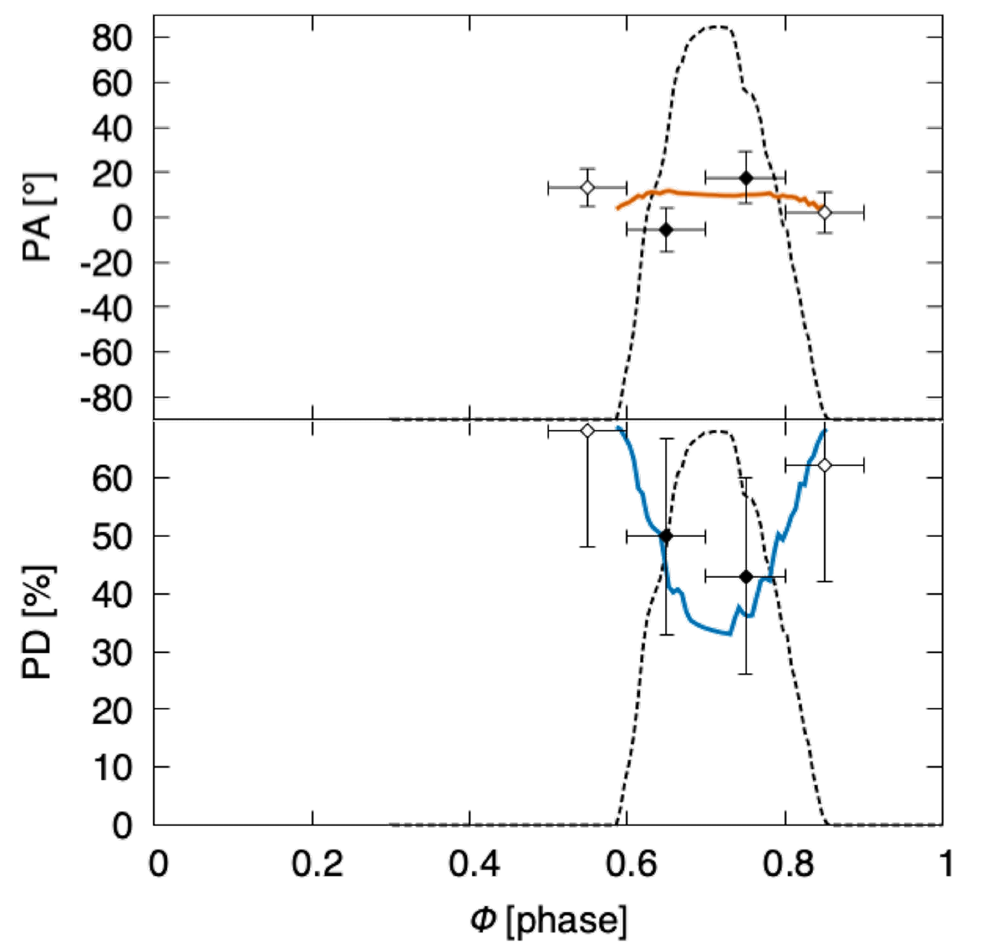}
 \caption{Same figure with Fig. \ref{fig:b1509}. The black points show the observed PA and PD data for PSR B0540-69 \citep{2024ApJ...962...92X}, where the hollow points correspond to the phases with low calculated intensity. 
 The other parameters are $\alpha = 10^{\circ}$, $a = 0.95$, $\sin\theta_{\rm p}(R_{\rm LC})=0.06$, $\xi = 110^{\circ}$.}
 \label{fig:b0540}  
\end{figure}

\citet{2024ApJ...962...92X} reported polarization characteristics of PSR B0540-69 and showed PD of $68.1\pm20.2\%$ and $62.4\pm20.1\%$ at the two edge phase bins (width 0.1 phase) in the broad intensity peak (width $\sim0.3$ phase), and lower PD ($49.5\pm16.6\%$ and $43\pm16\%$) between these bins in the $4-6 ~{\rm keV}$ X-ray band.
The PA changes about $23.0^{\circ}$ between the 4 bins of the intensity peak.
In the outer gap model, the outward emission could reproduce this single broad intensity peak with large PD of near $30\%-70\%$ as in Fig. \ref{fig:b0540} with $\alpha = 10^{\circ}$, $a = 0.95$, $\sin\theta_{\rm p}(R_{\rm LC})=0.06$, $\xi = 110^{\circ}$.
The model reference phase was shifted by 0.3 to align the X-ray intensity peaks shown in Fig. \ref{fig:b0540}. In addition, the PA offset of $-46^{\circ}$ was applied to the model PA in Fig. \ref{fig:b0540}. Since the photon count at the intensity peak fell below the threshold at the same resolution used previously (Fig. \ref{fig:results_outward_dependence_alla}, \ref{fig:results_inward_dependence_alla}, \ref{fig:inward_dependence_innerb}-\ref{fig:results_overlap}), we performed the calculation by quadrupling the number of magnetic field lines (to 4096).
A small variation of PA in Fig. \ref{fig:b0540} is not contradict to the observed PA with errors significantly.
Thus, observed high PD indicates that outward emission is dominant for the PSR B0540-69.
Finally, \citet{2025A&A...699A..33B} have reported the PD of $\lesssim 50\%$ for PSR J0205+6449.
Our results of both outward and inward emissions which have PD of under $20\%$ at the intensity peaks (Table. \ref{tab:table01}) are consistent with this observed data. 

The phase-resolved X-ray polarization of the Vela pulsar has not been observed so far although the phase-average X-ray PD is thought to be low ($\lesssim 8 \%$) \citep{2022Natur.612..658X}.
The Vela pulsar shows X-ray intensity peaks in phases 0.50, 0.90, and 0.95, where no gamma-ray peaks are observed, and it is difficult to explain these X-ray peaks by the outward emission in the outer gap model \citep{2008MNRAS.386..748T}.
\citet{2008MNRAS.386..748T} have suggested that its characteristics could be explained by considering the inward emission (see also \citet{2011ApJ...739...14K}).
The polarization properties of these inward emission are calculated (Figs. \ref{fig:results_inward_dependence_alla}, \ref{fig:results10_inner_boundary}, and \ref{fig:inward_dependence_innerb}).
For $\alpha=65^{\circ}$, $a=0.95$, and $\xi=100^{\circ}$, the inward EP at 0.50 phase has the PD of $\simeq5\%$ (Fig. \ref{fig:results_inward_dependence_alla} (A)), the inward LP at 0.90 phase has the PD of $\simeq10\%$ (Fig. \ref{fig:results_inward_dependence_alla} (B)), and the inward LP2 at 0.95 phase has the PD of $\simeq1\%$ (Fig. \ref{fig:inward_dependence_innerb}).
It could be an important clue to demonstrate the inward emission and the outer gap model if the PD values at each intensity peaks of the Vela pulsar are observed.
To further constrain the emission model considering inward emission, more phase-resolved polarization observation and that of many pulsars 
are important.
IXPE and future polarimeter satellites such as eXTP \citep{2025SCPMA..6819502Z} will serve these observation data.

\acknowledgments
We are grateful to the anonymous referee for constructive comments.
This work was supported by JST SPRING, Grant Number JPMJSP2132 (S.S.).
This work was also supported by JSPS Grants-in-Aid for Scientific Research 22K03681, 23K22538, 23K20038 and 26K07066 (S.K.).
J.T. are supported by the National Key Research and Development Program of China (grant No. 2020YFC2201400) and the National Natural Science Foundation of China (grant No. 12573044).

\bibliographystyle{apj_8}
\bibliography{ref}

\end{document}